\documentclass[fleqn,10pt]{wlscirep}
\usepackage[utf8]{inputenc}
\usepackage[T1]{fontenc}
\usepackage{hyperref}
\usepackage{booktabs}
\usepackage{pythonhighlight}
\usepackage{svg}
\newcommand{\gcheck}{{\color[RGB]{0, 165, 100}\checkmark}}
\usepackage{color, colortbl}
\usepackage{xcolor}
\definecolor{Gray}{gray}{0.9}
\usepackage{subcaption}
\usepackage{dirtree}
\usepackage[draft]{minted}
\newenvironment{code}{\captionsetup{type=listing}}{}
\usepackage{caption}
\usepackage{newfloat}

\title{\textit{pyCANON}: A Python library to check the level of anonymity of a dataset} 

\author[1,$\dag$,*]{Judith Sáinz-Pardo Díaz}
\author[1,$\dag$]{Álvaro López García}
\affil[1]{Instituto de Física de Cantabria (IFCA), CSIC-UC \\ Avda. los Castros s/n. 39005 - Santander (Spain)}

\affil[*]{corresponding author: Judith Sáinz-Pardo Díaz (sainzpardo@ifca.unican.es)}

\affil[$\dag$]{these authors contributed equally to this work}

\begin{abstract}

Openly sharing data with sensitive attributes and privacy restrictions is a challenging task. In this document we present the implementation of \textit{pyCANON}, a Python library and command line interface (CLI) to check and assess the level of anonymity of a dataset through some of the most common anonymization techniques: \textit{k-anonymity}, \textit{($\alpha$,k)-anonymity}, \textit{$\ell$-diversity}, \textit{entropy $\ell$-diversity}, \textit{recursive (c,$\ell$)-diversity}, \textit{basic $\beta$-likeness}, \textit{enhanced $\beta$-likeness}, \textit{t-closeness} and \textit{$\delta$-disclosure privacy}. For the case of more than one sensitive attributes, two approaches are proposed for evaluating this techniques. The main strength of this library is to obtain a full report of the parameters that are fulfilled for each of the techniques mentioned above, with the unique requirement of the set of quasi-identifiers and that of sensitive attributes. We present the methods implemented together with the attacks they prevent, the description of the library, use examples of the different functions, as well as the impact and the possible applications that can be developed. Finally, some possible aspects to be incorporated in future updates are proposed.

\end{abstract}
\begin{document}

\flushbottom
\maketitle
\thispagestyle{empty}

\section*{Introduction}\label{sec:introduction}

The unstoppable advances in data analysis and processing techniques for knowledge extraction and decision making, whether concerning Big Data or small data, motivate the idea of publishing datasets in an accessible way for the scientific community and society in general. In the same way, the need for collaboration between different institutions, research centers, or even companies, means that they need to be able to share the necessary data among themselves in a secure way. Furthermore, the need to publish open data in order to build an informed society about, for example, the processes of public organizations, intensifies the need to develop tools that allow the publication of such data with privacy guarantees.

When discussing the privacy guarantees that a dataset must fulfill, we are referring to preventing an attacker from extracting sensitive information from a specific individual. There is a strong requirement to protect certain data, such as health records (hospital visits, chronic illnesses), banking information or even police reports. For example, in the case of mental health records, the disclosure of such data about an individual could lead to discrimination and social exclusion. A person's medical data can also be used in an unethical manner to increase the cost of health insurance, or even lead to social foreclosure if it is revealed that a certain individual suffers from a specific disease. Similarly, information on political beliefs or even on an individual's monthly income could lead to exclusion and prejudice. Every day we are all producing a digital footprint that reveals our tastes, hobbies or even fears. In many cases it is reflected in personalized advertising, but to what extent is this good for the user? If someone always consults news of certain political behavior, some algorithms will hide news that show contrary opinions, resulting in an intellectual isolation of the users and a lack of critical thinking of the society. 

A classic example of data that was openly published and eventually showed a security breach is that of the US Census. Specifically, a study conducted on 1990 U.S. Census data revealed that 87$\%$ of the time, three pieces of information (zip code, gender and date of birth) were sufficient to identify someone in a database (see \cite{sweeney2000simple}). This shows that it is not enough to remove identifiers from a database to prevent an individual from being identified, since he/she can be identified by other types of attributes which a priori seem harmless. Let's briefly define four key concepts about the attributes or columns of a database. Firstly, the \textbf{\textit{identifiers}} are those variables in a database that allow to identify an individual (e.g. name, ID number, email...). On the other hand, the \textbf{\textit{quasi-identifiers (QI)}} are variables that although a priori seem not to show relevant information, combined between them can make possible the identification of an individual, as occurs in the case of the U.S. Census. Moreover, they are accessible to the attacker (e.g. gender, age, city, etc.). Finally, we have the \textbf{\textit{sensitive attributes (SA)}}, which are the columns of the database that contain sensitive information that must not be disclosed and must not be able to be extracted. 

To this aim, several software projects and libraries emerge. The main tool to highlight when talking about data anonymization is the ARX Software\cite{prasser2015putting}: ARX is a comprehensive open source software for anonymization of sensitive data that supports a wide variety of privacy, risk and quality models, together with methods for data transformation and for analyzing the usefulness of the output data. PRIVAaaS \cite{7973820} is another tool focused on providing policy-based anonymization in distributed data processing environments, aimed at reducing the data leakage in Big Data processing. ARGUS (including the $\mu$-ARGUS and $\tau$-ARGUS packages) is a software library for Statistical Disclosure Control delivered by the CASC-project focused on microdata and tabular data.


This paper presents \textit{pyCANON}, a Python library and CLI that allows to check if certain anonymity conditions are verified for a dataset, given a set of quasi-identifiers and a set of sensitive attributes. The main idea behind it is to provide researchers, and in general anyone who wants to publish a dataset in open access or to share it with others, with a prior knowledge of the level of anonymization of their data. This will provide insights about the possible risks to which these information would be exposed, allowing to verify the impact and their resistance to different attacks. An important consideration is that these data have to be provided in tabular format, that is, they cannot be in images, videos or any other kind of unstructured data.

In the following sections, the theoretical bases of the implemented methods are presented, as well as a brief explanation of the main functionalities of the software, examples of use, and the impact and different applications associated with it.

\section*{Results}\label{sec:results}

As explained in the previous section, the purpose of the library presented in this article is, essentially, to check the anonymity level of a dataset. For this purpose, we propose the use of nine different anonymization techniques: \textit{k-anonymity}, \textit{($\alpha$,k)-anonymity}, \textit{$\ell$-diversity}, \textit{entropy $\ell$-diversity}, \textit{recursive (c,$\ell$)-diversity}, \textit{basic $\beta$-likeness}, \textit{enhanced $\beta$-likeness}, \textit{t-closeness} and \textit{$\delta$-disclosure privacy}. Specifically, given a set of data, a list of quasi-identifiers and a list of sensitive attributes, it will be possible to check for which parameters each of the aforementioned techniques are verified, in order to know the degree of anonymity of such data, and thus the possible risks they may be subject to. 

Before exposing the definitions and different aspects taken into account for the implementation of the different techniques, let us present the concept of \textbf{\textit{equivalence class (EC)}}. An equivalence class is a partition of a database in which all the quasi-identifiers have the same value. That is, users who are in the same equivalence class are all indistinguishable with respect to the quasi-identifiers. The different techniques under study are presented below:

\begin{itemize}
    \item \textbf{\textit{k-anonymity.}} A database verifies \textit{k-anonymity} if each equivalence class of the database has at least $k$ rows. In other words, for each row of the database, there are at least $k-1$ indistinguishable rows with respect to the quasi-identifiers. Note that $k\geq1$ is always verified.
    \item \textbf{\textit{($\alpha$,k)-anonymity.}} Given only one sensitive attribute $S$, it is checked if the database is \textit{k-anonymous} and the frequency of each possible values of $S$ is lower or equal than $\alpha$ in each equivalence class. 
    \item \textbf{\textit{$\ell$-diversity.}} In the case of a single sensitive attribute $S$, it is satisfied if for each equivalence class, there are at least $\ell$ distinct values for $S$. Note that $\ell \geq1$ is always verified.
    \item \textbf{\textit{Entropy $\ell$-diversity.}} A database with a single sensitive attribute $S$ verifies this condition if $H(EC)>log(\ell)$, for every equivalence class \textit{EC} of the database. Note that $H(EC)$ is the entropy of the equivalence class \textit{EC}, defined as: 
    
    $$
    H(EC) = -\sum_{s\in D} p(EC, s)\log(p(EC, s)), 
    $$
    with $D$ the domain of $S$, and $p(EC,s)$ the fraction of records in $EC$ that have $s$ as sensitive attribute.
    
    \item \textbf{\textit{Recursive (c,$\ell$)-diversity.}} The main potential of this technique is that if a value of the sensitive attribute $S$ is removed from an equivalence class which verifies \textit{(c,$\ell$)-diversity}, then \textit{(c,$\ell$-1)-diversity} is preserved. For the implementation of this technique, \cite{1617392} has been used as reference (in order to get the formal definition of the concept). Specifically, suppose there are $n$ different values for a sensitive attribute $S$ in an equivalence class $EC$. Be $r_{i}$ $(i\in\{1,\hdots,n\})$ the number of times that the i-th most frequent value of $S$ appears in $EC$. Then, $EC$ verifies \textit{recursive (c,$\ell$)-diversity} for $S$ if $r_{1} < c(r_{l}+r_{l+1}+...+r_{n})$. In view of the previous inequality, in our implementation of this technique the value of $c$ will not be calculated if a value of $\ell=1$ is obtained.
    
    \item \textbf{\textit{Basic $\beta$-likeness and enhanced $\beta$-likeness.}} These two techniques have been implemented following the definitions 2 and 3 of \cite{cao2012publishing}, and can be used in order to control the distance between the distribution of a sensitive attribute in an equivalence class and in the entire database. In particular, be $\mathcal{P}=\{p_1,\hdots,p_n\}$ the distribution of a sensitive attribute $S$ in the whole database and $\mathcal{Q}=\{q_1,\hdots,q_n\}$ that of an equivalence class $EC$. Be $max(D(\mathcal{P},\mathcal{Q}))=max\{D(p_{i},q_{i}): p_{i}\in \mathcal{P}, q_{i}\in \mathcal{Q} \land p_{i}<q_{i}\}$, then \textit{basic $\beta$-likeness} is verified if $max(D(\mathcal{P},\mathcal{Q})) \leq \beta$ and \textit{enhanced $\beta$-likeness} is verified if $D(p_{i},q_{i}) \leq min\{\beta, -\log(p_{i})\}$ $\forall q_{i}\in \mathcal{Q}$. In both cases $\beta > 0$.
    Note that \textit{enhanced $\beta$-likeness} provide more robust privacy than \textit{basic $\beta$-likeness}.
    
    In our implementation the relative distance function is considered in order to calculate the distance between the distributions, that is: $D(p_{i},q_{i}) = \frac{q_{i}-p_{i}}{p_{i}}$.
    
    \item \textbf{\textit{t-closeness.}} The goal is again similar of that of the two previous techniques. A database with one sensitive attribute $S$ verifies \textit{t-closeness} if all the equivalence classes verify it. An equivalence class verify \textit{t-closeness} if the distribution of the values of $S$ are at a distance no closer than $t$ from the distribution of the sensitive attribute in the whole database. In order to measure the distance between the distributions, following \cite{4221659}, the Earth Mover’s distance (EMD) between the two distributions using the ordered distance is applied for numerical sensitive attributes. For categorical attributes, the equal distance is used.
    
    \item \textbf{\textit{$\delta$-disclosure privacy.}} Be a database with only one sensitive attribute $S$, $p(EC,s)$ the fraction of records with $s$ as sensitive attribute in the equivalence class EC, $p(DB,s)$ that for the whole database (DB). Then, \textit{$\delta$-disclosure privacy} is verified iff:
    $$
    \left|\log\left(\frac{p(EC,s)}{p(DB,s)}\right)\right|< \delta,
    $$ 
    for every $s\in D$ (with $D$ the domain of $S$) and every equivalence class \textit{EC} \cite{10.1145/1401890.1401904}.
\end{itemize}

The motivation for including the nine techniques outlined above, and not a smaller number of them or just the most classic ones (e.g. \textit{k-anonymity} or \textit{$\ell$-diversity}), is that they are not all useful against the same types of attacks. That is, there are techniques that are very useful against certain attacks, but cannot prevent from others. In particular, Table~\ref{tab:attacks} briefly describes some of the most common attacks that databases can suffer, namely: \textit{linkage}, \textit{re-identification}, \textit{homogeneity}, \textit{background knowledge}, \textit{skewness}, \textit{similarity} and \textit{inference attacks}. 

\begin{table}[ht]
    \centering
    \begin{tabular}{rl}
    \toprule
         \textbf{Attack} & \textbf{Description}  \\
         \midrule
         \rowcolor{Gray}\textit{Linkage} & Consists of combining at least two anonymized databases in order to reveal the identity of\\
         \rowcolor{Gray}& some individuals present in both.\\
         \textit{Re-identification} & This kind of attacks occurs when the anonymization process is reversed.\\
         \rowcolor{Gray}\textit{Homogeneity} & Can occur when all the values for a sensitive attribute in an equivalence class are identical.\\
         \textit{Background knowledge} & In this case, the adversary has some foreknowledge about the target of the attack e.g. knows\\
         & some auxiliary information about an individual in the database).\\
         \rowcolor{Gray}\textit{Skewness} & Can be carried out when there is an unfrequent value for a sensitive attribute in the whole \\
         \rowcolor{Gray}& database which is extremely frequent in an equivalence class.\\
         \textit{Similarity} & May occur when the values of a sensitive attribute in an equivalence class are semantically\\
         & similar (although different).\\
         \rowcolor{Gray}\textit{Inference} & Consists of using data mining techniques in order to extract information from the data.\\
    \bottomrule
    \end{tabular}
    \caption{Common attacks on databases and description.}
    \label{tab:attacks}
\end{table}
In addition, Table~\ref{tab:techniques_attacks} shows the most convenient techniques (although not the only ones) that can be applied to prevent each of the previously mentioned attacks.

\begin{table}[ht]
    \centering
    \resizebox{\linewidth}{!} {
    \begin{tabular}{rccccccc}
    \toprule
         \multicolumn{1}{c}{ } &  \multicolumn{7}{c}{\textbf{Principal attack which prevents}} \\
         \cmidrule{2-8}
         \textit{Technique} &  \textbf{Linkage} & \textbf{Re-identification} & \textbf{Homogeneity} & \textbf{Background} & \textbf{Skewness} & \textbf{Similarity} & \textbf{Inference}\\
         \midrule
         \textit{k-anonymity} & \gcheck & \gcheck & & & & &\\
         \textit{($\alpha$,k)-anonymity} & \gcheck & \gcheck & \gcheck & &  & &\\
         \textit{$\ell$-diversity} &  &  & \gcheck & \gcheck & & &\\
         \textit{Entropy $\ell$-diversity} &  &  & \gcheck & \gcheck & & & \\
         \textit{Recursive (c,$\ell$)-diversity} &  &  & \gcheck & \gcheck & & & \\
         \textit{t-closeness} &  &  &  &  & \gcheck & \gcheck & \\
         \textit{Basic $\beta$-likeness} &  &  &  &  & \gcheck & & \\
         \textit{Enhanced $\beta$-likeness} &  &  &  &  & \gcheck & & \\
         \textit{$\delta$-disclosure privacy} &  &  &  &  & \gcheck & & \gcheck\\ 
         \bottomrule
    \end{tabular}}
    \caption{Anonymization techniques and principal attacks that prevent (among others).}
    \label{tab:techniques_attacks}
\end{table}

It is important to take into account that the values of $t$ and $\delta$ for \textit{t-closeness} and \textit{$\delta$-disclosure privacy} respectively must be strictly greater than the ones obtained using \textit{pyCANON} (see the definition of that techniques).

On the other hand, it should be noted that although the anonymization techniques have been presented for the case where the database consists of a single sensitive attribute, they can nevertheless be applied in the case of multiple sensitive attributes. The latter may be quite common since in many use cases there are several attributed deemed to be sensitive information. Specifically, this library implements two approaches that can be followed in this case:
\begin{enumerate}
    \item In the simplest case, generalization is applied to the case of multiple sensitive attributes. That is, for each sensitive attribute (SA) each of the properties is checked, and the parameter that is satisfied for all of them is kept (e.g. for \textit{$\ell$-diversity} the smallest value of $\ell$ once it is computed for each SA will be kept, while the value of $\alpha$ for \textit{($\alpha$,k)-anonymity} will be the largest value of $\alpha$ of those obtained for each SA). In the following we will refer to this approach as the \textit{generalization} one.
    \item In the second approach, we will have to update the set of quasi-identifiers according to the sensitive attribute to be analyzed, and proceed again as in the previous case. That is, be $Q$ initial the set of quasi-identifiers, and $S$ the set of sensitive attributes. For each sensitive attribute $S_{i} \in S$, the set of quasi-identifiers considered in each case would change (and with it the different equivalence classes), so that it will be $Q \cup (S \setminus S_{i})$. 
    In the following we will refer to this approach as that of the \textit{quasi-identifiers update}. The idea of introducing this second approach is that in certain cases an attacker could know some of the sensitive attributes, thus acting as quasi-identifiers that would allow inferring information about the rest of the sensitive information.
\end{enumerate}

As will be explained in the following, \textit{pyCANON} implements the two previously exposed approaches, and it is up to the user to select which one to use. One of the main points to take into account is that the second approach, in which the set of quasi-identifiers is updated according to the sensitive attribute to be analyzed, is more computationally expensive, since for each sensitive attribute the equivalence classes must be recalculated (because the set of quasi-identifiers varies, in addition to being larger). The different functions available on this framework are shown in the following section, together with some use examples.

\section*{Discussion}
\label{sec:Discussion}
As already stated, the purpose of our tool is to assess the level of anonymity of a dataset with regards to the most common techniques, being complementary to other anonymization tools like ARX. In this section, a battery of applications that validate the impact and usefulness of this library for different purposes is discussed. In particular, different examples are presented using openly available data, so that the reproducibility of the results is ensured. However, note that as will be explained later, many of these data have been anonymized using the open source software \textit{ARX}\cite{prasser2015putting}, establishing different levels of hierarchies for the generalization of the quasi-identifiers, which could cause the results to vary according to the different hierarchies introduced.

Suppose we have a dataset and a list of quasi-identifiers for which we want to get the data to verify \textit{k-anonymity} with, for instance, $k = 5$. To do this, we can use a software like \textit{ARX}, selecting as quasi-identifiers the columns of the data set that interest us. Suppose that by mistake, one of the quasi-identifiers is introduced as an insensitive attribute in the anonymization process (for example, in \textit{ARX} there is a drop-down to distinguish the type of attribute, so it is not complex to imagine a possible human error). Once anonymized using this software, we obtain a dataset verifying \textit{k-anonymity} with $k=5$ for the columns included as quasi-identifiers. Then, \textit{pyCANON} can be used to check if the new dataset truly verifies \textit{k-anonymity} for $k=5$ for the initial given list of quasi-identifiers. However, suppose that by doing this check we get a value $k=1$, i.e., there is at least one set of quasi-identifiers which only has a single individual (an equivalence class with only one row). This allows us to quickly detect that there has been a failure in the anonymization process (in this case, at least one of the columns that should be a QI, was introduced by mistake in \textit{ARX} as insensitive value). In this way, by checking the anonymity level with \textit{pyCANON} we have managed to prevent a bug that can be very common, but that would allow an attacker to extract unwanted information about the dataset. 

Another example of using the library will be now presented with the stroke dataset \cite{stroke_data} with the following set of quasi-identifiers (QI) and sensitive attributes (SA): \textbf{QI = [`gender', `age', `hypertension', `heart\_disease', `ever\_married', `work\_type', `Residence\_type', `smoking\_status']} and \textbf{SA = [`stroke']}. Again, using \textit{ARX} \textit{k-anonymity} will be applied for different values of $k$, namely $k=2, 5, 10, 20, 25$. However, for example by setting $k = 20$ we can check with \textit{pyCANON} that the actual $k$ value obtained for the resulting dataset is $k=22$. This does not mean that the anonymization done with \textit{ARX} is wrong, because if \textit{k-anonymity} is verified for $k=22$, it is evidently verified for $k=20$. In fact the software itself warns us of this if we enter the data, where it indicates that the size of the smallest equivalence class is 22. This is probably because with the generalization allowed, the greater value of $k$ which verifies \textit{k-anonymity} with $k=20$, is 22. Similarly, if in this same example we set $k=25$ in \textit{ARX}, then we can check with \textit{pyCANON} that the maximum value for $k$ is 27.

Another practical example is the following: suppose we have a dataset to which \textit{k-anonymity} has been applied for a certain value of $k$, again, for example, using \textit{ARX}. We may be interested, without applying any further technique, to check some of the other previously exposed anonymization techniques, and compare how these scale as a function of the value of $k$. For example, given the drugs dataset \cite{drug_data} with \textbf{QI = [`Age', `Sex', `BP', `Cholesterol', `Na\_to\_K']} and \textbf{SA = [`Drug']}, anonymized using \textit{ARX} only considering \textit{k-anonymity} for different values of $k$, namely $k=2, 5, 10, 15, 20$. Let's see in Table \ref{tab:table_k} how scale, for example, the values of $\ell$ for \textit{$\ell$-diversity}, \textit{t} for \textit{t-closeness}, and $\beta$ for \textit{basic $\beta$-likeness} (without removing the values suppressed when anonymizing with \textit{ARX}).

\begin{table}[ht]
    \centering
    \begin{tabular}{cccc}
    \toprule
     \textbf{\textit{k (k-anonymity)}} & \textit{$\ell$ ($\ell$-diversity)} & \textit{t (t-closeness)} &\textit{ $\beta$ (basic $\beta$-likeness)}\\
     \midrule
     \textbf{2} & 1 & 0.93 & 11.5\\
     \textbf{5} & 2 & 0.68 & 8.38\\
     \textbf{10} & 2 & 0.63 & 5.62\\
     \textbf{15}* & 2 & 0.48 & 5.62\\
     \textbf{20}** & 2 & 0.38 & 2.32\\
     \bottomrule
    \end{tabular}
    \caption{Values of $\ell$ for \textit{$\ell$-diversity}, \textit{t} for \textit{t-closeness}, and $\beta$ for \textit{basic $\beta$-likeness} obtained for a prefixed value of $k$ for \textit{k-anonymity} considering the drug dataset.\\
    * $k=15$ has been set with ARX, and can be checked with \textit{pyCANON} that the resulting data also verify \textit{k-anonymity} for $k=17$.\\
    ** $k=20$ has been set with ARX, and can be checked with \textit{pyCANON} that the resulting data also verify \textit{k-anonymity} for $k=21$.}
    \label{tab:table_k}
\end{table} 

Thus, as can be seen in Table~\ref{tab:table_k}, it is enough to apply the \textit{k-anonymity} process (which requires little computational cost and, depending on the size of the dataset, in \textit{ARX} is done in a few seconds), to obtain better values for the parameters of other anonymization techniques, such as increasing the value of $\ell$ for \textit{$\ell$-diversity}, or decreasing the value of $t$ for \textit{t-closeness}. It can also be applied to a dataset that has not been previously processed, to check which techniques should be applied, which are the most significant quasi-identifiers (that provide more information), etc. For example, given an unprocessed dataset, the values of $k$ and $\ell$ obtained with the different permutations of the quasi-identifiers could be calculated, in order to evaluate the possibility of eliminating some of them (cataloging them as sensitive attributes for instance). 

\begin{figure}[ht]
    \centering
    \includegraphics[width = 0.55\textwidth]{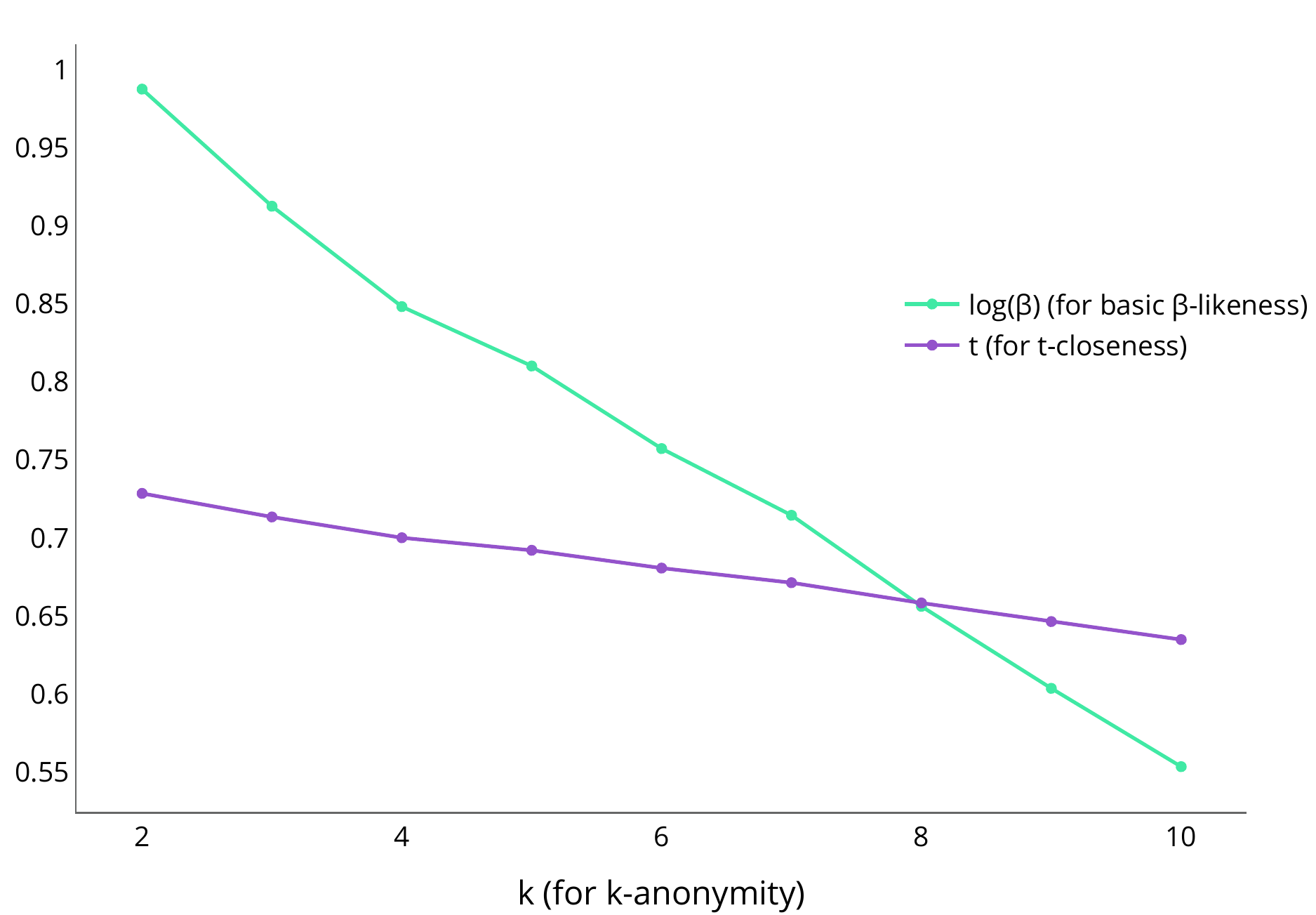}
    \caption{Evolution of $t$ for \textit{t-closeness} and $\log(\beta)$ for \textit{basic $\beta$-likeness} when varying the value of $k$ for \textit{k-anonymity}.}
    \label{fig:k_vs_betat}
\end{figure}

The same has been done again for the adult database, with the quasi-identifiers and sensitive-attributes given in the first example of the \textit{Usage Examples} subsection. Specifically, the dataset has been anonymized using \textit{ARX} considering all $k\in [2, 10]\cap \mathbb{N}$ for \textit{k-anonymity}. Figure \ref{fig:k_vs_betat} shows the evolution of the values of $t$ and $\beta$ (the latter is shown in logarithmic scale for better visualization) for \textit{t-closeness} and \textit{basic $\beta$-likeness} respectively, removing the values suppressed by \textit{ARX}. A substantial decrease is produced both for the values of $t$ a $\beta$ (remember that $t$ is strictly greater than the value obtained using \textit{pyCANON}). 

Finally, let us see the results obtained with \textit{pyCANON} for three different techniques of the nine described above after anonymizing using the ARX Software. The adult dataset will be used with the quasi-identifiers and sensitive attribute presented in the previous examples. With this example we want to show the usefulness of this library to check that an anonymity process has been carried out correctly. Tables~\ref{tab:t_closs} and \ref{tab:delta_disc} show the results obtained by setting certain values of $t$ and $\delta$ for \textit{t-closeness} and \textit{$\delta$-disclosure} respectively in \textit{ARX}, and that obtained using \textit{pyCANON} (without removing the records suppressed when anonymizing with \textit{ARX}). 

\begin{table}[ht]
\parbox{.45\linewidth}{
\centering
\begin{tabular}{c|c}
    \toprule
         \textit{t for t-closeness} & \textit{t for t-closeness}\\
         \textbf{\textit{ARX}} & \textbf{\textit{pyCANON}}\\
         \midrule
         0.1 & 0.09795 \\
         0.05 & 0.04773 \\
         0.04 & 0.03484 \\
         0.03 & 0.02828 \\
         0.02 & 0.01550 \\
         0.01 & 0.00919\\
    \bottomrule
    \end{tabular}
    \caption{Values for \textit{t-closeness} using \textit{ARX} and \textit{pyCANON}.}
    \label{tab:t_closs}
}
\hfill
\parbox{.45\linewidth}{
\centering
    \begin{tabular}{c|c}
    \toprule
         \textit{$\delta$ for $\delta$-disclosure} & \textit{$\delta$ for $\delta$-disclosure}\\
         \textbf{\textit{ARX}} & \textbf{\textit{pyCANON}}\\
         \midrule
         0.1 &  0.06237 \\
         0.05 & 0.03339 \\
         0.04 & 0.01661 \\
         0.03 & 0.01661 \\
         0.02 & 0.01352 \\
         0.01 & 0.00293 \\
    \bottomrule
    \end{tabular}
    \caption{Values for \textit{$\delta$-disclosure} using \textit{ARX} and \textit{pyCANON}.}
    \label{tab:delta_disc}
}
\end{table}

It should be noted that in the case of \textit{basic $\beta$-likeness}, differences may occur depending on the function used to calculate the distance. In the case of \textit{pyCANON}, the relative distance is implemented. However, Table~\ref{tab:basic_beta} shows five examples where the values established with \textit{ARX} and the results with \textit{pyCANON} are compared. The same will occur with other methods depending on the strategy or definition considered in each case, as they may differ.
\begin{table}[ht]
    \centering
    \begin{tabular}{c|c}
    \toprule
         \textit{$\beta$ for basic $\beta$-likeness} & \textit{$\beta$ for basic $\beta$-likeness}\\
                  \textbf{\textit{ARX}} & \textbf{\textit{pyCANON}}\\
         \midrule
         0.1 & 0.09766 \\
         0.05 & 0.04909 \\
         0.01 & 0.00523 \\
         0.005 & 0.00425 \\
         0.001 & 0.00093 \\
    \bottomrule
    \end{tabular}
    \caption{Values for \textit{basic $\beta$-likeness} using \textit{ARX} and \textit{pyCANON}.}
    \label{tab:basic_beta}
\end{table}

In addition, several test have been carried out for the value of $k$ of \textit{k-anonymity}, using five different datasets. All of then can be found in the folder \textit{examples} of the original framework repository. In these examples, a case with several sensitive attributes is included, and the values obtained with each of the two proposed approaches are studied.

All these functionalities will allow researchers and users in general to improve their practices regarding the publication of data using anonymization techniques, thus helping to raise awareness of the importance of the anonymization process. That is, it will enable data to be published with greater security guarantees. Again, if a database has been anonymized with a particular software, it would not be very appropriate to check the anonymization level with it, because if there is a bug, it would not be detected when checking. However, \textit{pyCANON} is a library that allows to check the correct application of these techniques, as well as to test for which parameters those techniques that have not been consciously applied would verify with. This would change the daily practices of those who are in charge of publishing data, but also for those institutions, centers or researchers who collaborate by sharing data between them, because they could be sure of the privacy level of the information which they are going to share.

It should be noted that this library does not require strong knowledge of Python language, since it is designed so that the user only has to enter a \textit{pandas} dataframe with the data (or even the path to the file where the data is stored in the case of using the CLI), a string list with the names of the columns that are quasi-identifiers, and another with those that are sensitive attributes. This makes it a library accessible to the general public wishing to check the anonymity level of a dataset regardless of its background. The only condition is that the data must have one of the following extensions: \textit{.csv}, \textit{.xls}, \textit{.txt}, or \textit{.sav}, or be stored in a \textit{pandas} dataframe. 

\section*{Methods}
\label{sec:methods}

In this library, the user is provided with a set of functions to calculate the anonymity level of a dataset based on the techniques mentioned previously. In addition, for the case of multiple sensitive attributes, it allows to know the results for the two approaches previously exposed, namely: \textit{generalization} or \textit{quasi-identifiers update}.

Specifically, a function for the calculation of each of the anonymity techniques presented has been implemented in the package \textit{anonymity}. Be \textit{data} a \textit{pandas} dataframe containing the dataset under study. Be \textit{qi} the list of quasi-identifiers and \textit{sa} that of the sensitive attributes. The parameter \textit{gen} indicates whether to generalize for the case of multiple SA: If true (default) generalization approach is applied, if False, the set of QI is updated for each SA. Then, the different functions used to calculate the parameters for the previously mentioned properties are that exposed in Table~\ref{tab:functions}. Remember that the values of $t$ and $\delta$ for \textit{t-closeness} and \textit{$\delta$-disclosure privacy} must be strictly greater than the ones obtained using the functions of \textit{pyCANON}.

\begin{table}[ht]
    \centering
    \begin{tabular}{rl}
    \toprule
    \textbf{Parameter calculated} & \textbf{Function}\\
    \midrule
    $k$ for \textit{k-anonymity} &  \textit{k\_anonymity(data, qi)}\\
    $\alpha$ and $k$ for \textit{($\alpha$,k)-anonymity} & \textit{alpha\_k\_anonymity(data, qi, sa, gen)}\\
    $\ell$ for \textit{$\ell$-diversity} & \textit{l\_diversity(data, qi, sa, gen)}\\
    $\ell$ for \textit{entropy $\ell$-diversity} &  \textit{entropy\_l\_diversity(data, qi, sa, gen)}\\
    $c$ and $\ell$ for \textit{recursive (c,$\ell$)-diversity} & \textit{recursive\_c\_l\_diversity(data, qi, sa, gen)}\\
    $\beta$ for \textit{basic $\beta$-likeness} & \textit{basic\_beta\_likeness(data, qi, sa, gen)}\\
    $\beta$ for \textit{enhanced $\beta$-likeness} & \textit{enhanced\_beta\_likeness(data, qi, sa, gen)}\\
    $t$ for \textit{t-closeness} & \textit{t\_closeness(data, qi, sa, gen)}\\
    $\delta$ for \textit{$\delta$-disclosure privacy} & \textit{delta\_disclosure(data, qi, sa, gen)}\\
    \bottomrule
    \end{tabular}
    \caption{\textit{pyCANON} main functions to check anonymity properties of a dataset. }
    \label{tab:functions}
\end{table}

\begin{figure}[ht]
    \centering
    \includegraphics[width = 0.7\linewidth]{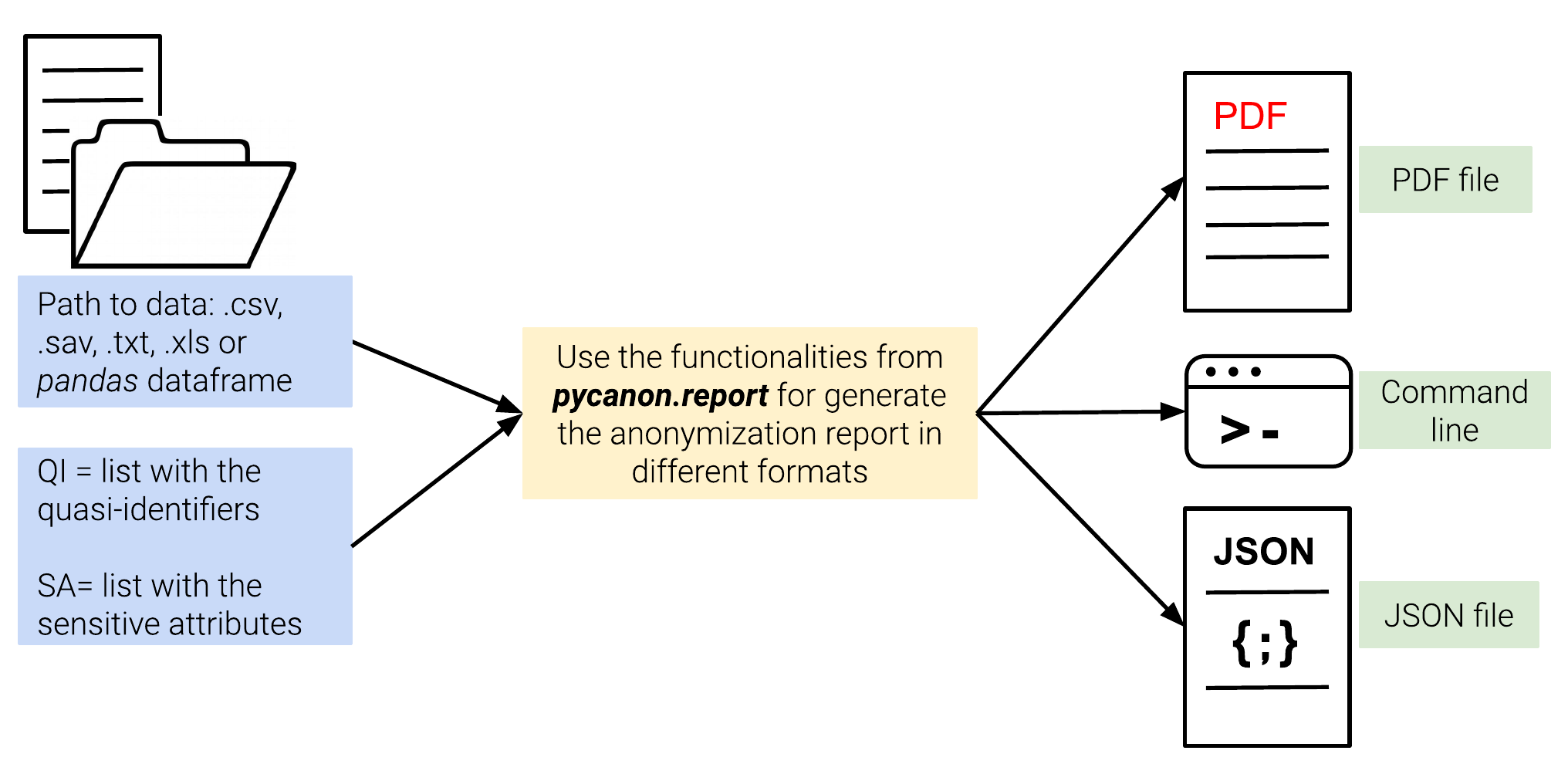} 
    \caption{Schema for obtaining the anonymization report from the data, list of quasi-identifiers and sensitive attributes.}
    \label{fig:flux_report}
\end{figure}

In addition, the package \textit{report} is a key functionality of this library. Again \textit{data}, \textit{qi}, \textit{sa} and \textit{gen} are defined as in the previous example. The purpose of this package is to generate a report with the anonymization level of the data file entered, checking all the techniques mentioned above. This report can be generated into a \textit{JSON} file, a \textit{PDF} file, or displaying it on the screen. An example will be shown in the following \textit{Usage examples} Subsection, but the basic schema using this package is represented in Figure \ref{fig:flux_report}.

Besides being a library with the two packages mentioned above: \textit{anonymity} and \textit{report}, \textit{pyCANON} is also implemented as a command line interface (CLI). The latter allows the user to execute the code via a simple command line with interactive use via text input.

Finally, it is important to note that the overall structure of the library includes, among others, unit tests, documentation, and files with example data as well as their corresponding tests. Also, note that the documentation of the library can be found at \url{https://readthedocs.org/projects/pycanon/}.

\subsection*{Usage examples}\label{sec:examples}
Let us start by presenting an example for a very classical dataset: adult dataset (see \cite{Dua:2019}). Specifically, the following quasi-identifiers (QI) and sensitive attributes (SA) have been chosen: \textbf{QI = [`age', `education', `occupation', `relationship', `sex', `native-country']} and \textbf{SA = [`salary-class']}. Be FILE\_NAME a string with the path to the \textit{.csv} file where the mentioned dataset is stored, suppose FILE\_NAME = `adult.csv'. First, we must import the package \textit{anonymity} from the \textit{pyCANON} library, and then run a function for each of the properties studied above, as follows: 

\begin{code}
\begin{minted}[fontsize=\small]{python}
import pandas as pd
from pycanon import anonymity

FILE_NAME = "adult.csv"
DATA = pd.read_csv(FILE_NAME)
QI = ["age", "education", "occupation", "relationship", "sex", "native-country"]
SA = ["salary-class"]

# Calculate k for k-anonymity:
k = anonymity.k_anonymity(DATA, QI)
# Calculate alpha for (alpha,k)-anonymity:
alpha, _ = anonymity.alpha_k_anonymity(DATA, QI, SA)
# Calculate l for l-diversity:
l_div = anonymity.l_diversity(DATA, QI, SA)
# Calculate l for entropy l-diversity:
entropy_l = anonymity.entropy_l_diversity(DATA, QI, SA)
# Calculate c for recursive (c,l)-diversity:
c, _ = anonymity.recursive_c_l_diversity(DATA, QI, SA)
# Calculate beta for basic beta-likeness:
basic_beta = anonymity.basic_beta_likeness(DATA, QI, SA)
# Calculate beta for enhanced beta-likeness:
enhanced_beta = anonymity.enhanced_beta_likeness(DATA, QI, SA)
# Calculate t for t-closeness:
t = anonymity.t_closeness(DATA, QI, SA)
# Calculate delta for delta-disclosure privacy:
delta = anonymity.delta_disclosure(DATA, QI, SA)
\end{minted}
\caption{Use example of the \textit{anonymity} package.}
\label{code:anonymity}
\end{code}\vspace{0.3cm}

The above example can be easily reproduced by simply installing the library, downloading the aforementioned dataset, reading it with pandas and applying the different functions from \textit{pycanon.anonymity}. 

Next, it is interesting to highlight how to proceed in the case of multiple sensitive attributes. The process is again very simple, but first the approach to be followed must be selected: \textit{generalization} or \textit{quasi-identifiers update}. In the first case, the procedure to follow is the same as the previous example, where the different functions were called for the case of a single sensitive attribute, but now, introducing two or more values in the list \textbf{SA}. By default the configuration of the functions selects the generalization approach if it detects more than one sensitive attribute. However, if one wants to use the approach of updating the quasi-identifiers (which is stricter from a privacy point of view), in all those functions where the list of sensitive attributes (\textbf{SA} in the example above) is entered, it is necessary to add a last attribute, \textit{gen = False}. This is indicating in each function that the conditions must be checked by updating the set of quasi-identifiers for each sensitive attribute under study. 

Regarding the process to obtain the anonymization report, the process to follow is simple. If one wants to save the report as a \textit{PDF} file, use the module \textit{pdf} from \textit{pycanon.report}, and just indicate the path to the PDF file where it is to be save in the variable $FILE\_PDF$, together with the data, quasi-identifiers, sensitive attributes and the value for the parameter \textit{gen}. In the same way, if one wants to save the report as a \textit{JSON} file, use the module \textit{json} from \textit{pycanon.report}. Note that it is also possible to display the report by command line using the function \textit{print\_report()} from the package \textit{report}. In the following, an example of displaying the report in a pdf file is shown.

\begin{code}
\begin{minted}[fontsize=\small]{python}
import pandas as pd
from pycanon.report import pdf

FILE_NAME = "adult.csv"
DATA = pd.read_csv(FILE_NAME)
QI = ["age", "education", "occupation", "relationship", "sex", "native-country"]
SA = ["salary-class"]
FILE_PDF = "report_adult.pdf"
pdf.get_pdf_report(DATA, QI, SA, file_pdf = FILE_PDF)
\end{minted}
\caption{Use example of the \textit{report} package}
\label{code:report}
\end{code}

\subsection*{Future work}
In the first release of this library, nine techniques exposed when presenting the methods have been included, but this software may be extended in future versions with new functionalities. Among them, the idea of applying techniques such as \textit{$\delta$-presence} or \textit{k-map}, for which it is necessary to use an auxiliary population, seems really attractive. In addition, it is also proposed for future updates and improvements of the library to include, together with the classification report that can be generated towards the \textit{report} package, personalized recommendations on the values obtained: for example, to inform that it is advisable to have a value of $\ell$ for \textit{$\ell$-diversity} strictly greater than 1 or that is recommended for the value of $\alpha$ for \textit{($\alpha$,k)-anonymity} to be strictly lower than 1. Furthermore, these recommendations could be customized according to the number of data, as well as quasi-identifiers and sensitive attributes involved.

\section*{Data availability}
No new data was generated in this work.

\section*{Code availability}
Code for the \textit{pyCANON} library and each compliance test is available in \url{https://github.com/IFCA/pycanon}.\\
The library documentation can be found in \url{https://pycanon.readthedocs.org}.

\section*{Acknowledgements} 

The authors acknowledge the funding through the European Commission - NextGenerationEU (Regulation EU 2020/2094), through CSIC's Global Health Platform (PTI Salud Global) and the support from the project AI4EOSC ``Artificial Intelligence for the European Open Science Cloud'' that has received funding from the European Union's Horizon Europe research and innovation programme under grant agreement number 101058593.

\section*{Author contributions statement}

Both authors conducted the research, designed, implemented and tested the library and wrote and review the manuscript.
\section*{Competing interests} 

The authors declare that they have no known competing financial interests or personal relationships that could have appeared to influence the work reported in this paper.

\end{document}